\documentstyle[prl,aps,multicol]{revtex} 

\newcommand{\bef}{\begin{figure}}
\newcommand{\enf}{\end{figure}}
\newcommand{\bec}{\begin{center}}
\newcommand{\enc}{\end{center}}

\def\TK{T_{\mbox{\scriptsize K}}}
\def\T2{T^{2}_{\mbox{\scriptsize K}}}
\def\Tpr{T^{\prime}_{\mbox{\scriptsize K}}}
\def\Tp2{T^{\prime 2}_{\mbox{\scriptsize K}}}
\def\nd{n_{\mbox{\scriptsize d}}}
\def\Vg{V_{\mbox{\scriptsize g}}}
\def\Vds{V_{\mbox{\scriptsize ds}}}
\def\Vac{V_{\mbox{\scriptsize ac}}}
\def\Tb{T_{\mbox{\scriptsize base}}}

\def\Lap{{\scriptscriptstyle\stackrel{<}{\sim}}}
\def\Gap{{\scriptscriptstyle\stackrel{>}{\sim}}}

\begin{document}
\bibliographystyle{prsty}
\draft
\title{From the Kondo Regime to the Mixed-Valence Regime in a
Single-Electron Transistor}
\author{D. Goldhaber-Gordon\cite{byline}, J. G\protect{\"{o}}res,
and M.A. Kastner} 
\address{
Department of Physics, Massachusetts Institute of Technology\\
Cambridge, MA 02139
}
\author{Hadas Shtrikman, D. Mahalu, and U. Meirav}
\address{
Braun Center for Submicron Research\\
Weizmann Institute of Science\\
Rehovot, Israel 76100
}
\date{\today}
\maketitle
\begin{abstract}
We demonstrate that the conductance through a single-electron
transistor at low temperature is in quantitative agreement with
predictions of the equilibrium Anderson model. When an unpaired
electron is localized within the transistor, the Kondo effect is
observed. Tuning the unpaired electron's energy toward the Fermi level
in nearby leads produces a cross-over between the Kondo and
mixed-valence regimes of the Anderson model.
\end{abstract}
\pacs{PACS 75.20.Hr, 73.23.Hk, 72.15.Qm, 73.23.-b}
\begin{multicols}{2}
\narrowtext

The effect of magnetic impurities on metals --- the Kondo effect ---
has been studied for half a century, and enjoys continued relevance
today in attempts to understand heavy-fermion materials and
high-$T_{\mbox{\protect\footnotesize c}}$ superconductors. Yet it has
not been possible to experimentally test the richly varied behavior
predicted theoretically. The theory depends on several parameters
whose values are not independently tunable for impurities in a metal,
and are often not even {\em a priori} known. On the other hand, it has
been predicted
\cite{nglee.kondo,glazmanraikh.kondo,meir.kondo,wingreen.kondo,izumida.kondo}
that a single-electron transistor (SET) should be described by the
Anderson impurity model, and hence should also exhibit the Kondo
effect. An SET contains a very small droplet of localized electrons,
analogous to an impurity, strongly coupled to conducting leads,
analogous to the host metal.  We have recently shown that when the
number of electrons in the droplet is odd, and hence one electron is
unpaired, the SET exhibits the Kondo effect\cite{gg.kondo} in
electronic transport. This observation has since been confirmed
\cite{delft.kondo,note.kondo} with additional quantitative detail. As
we show in this Letter, in SET experiments one can tune the important
parameters and test predictions of the Anderson model that cannot be
tested in bulk metals.

In our SET, a droplet of about $50$ electrons is
separated from two conducting leads by tunnel barriers. A set of
electrodes (Fig.\ \ref{energydiagram}(a)), on the surface of a
GaAs/AlGaAs heterostructure which contains a two-dimensional electron
gas (2DEG), is used to confine the electrons and create the tunnel
barriers. The 2DEG is depleted beneath the electrodes, and the narrow
constrictions between electrodes form the tunnel barriers.  Details
of the device fabrication and structure are given
elsewhere \cite{gg.kondo}.

In the Anderson model, the SET is approximated as a single localized
state, coupled by tunneling to two electron reservoirs. The state can
be occupied by $\nd = 0, 1,$ or $2$ electrons with opposite spin;
couplings to all other filled and empty states of the droplet are
neglected.  Adding the first electron takes an energy $\epsilon_0$
referenced to the Fermi level in the leads, but the second electron
requires $\epsilon_0 + U$, where the extra charging energy $U\
(1.9\pm0.05\, $meV in our SET) results from Coulomb repulsion. In the
diagram of Fig.\ \ref{energydiagram}(b), $\epsilon_0 < 0$, but
$\epsilon_0 + U > 0$, so there is one electron in the
orbital. However, this electron can tunnel into the leads, with rate
$\Gamma/h$, leading to Lorentzian broadening of the localized-state
energies with full width at half maximum (FWHM) $\Gamma$. The energy
$\epsilon_0$ can be raised by increasing the negative voltage $\Vg$ on
a nearby electrode (the middle left ``plunger gate'' electrode in
Fig.\ \ref{energydiagram}(a)) and $\Gamma$ can be tuned by adjusting
the voltages on the gates that form the constrictions.  Two other
important energies (not shown) are the spacing between quantized
single-particle levels $\Delta\epsilon \approx 400\,\mu$eV, and the
thermal broadening of the Fermi level in the leads $kT = 8 -
350\,\mu$eV.  The Kondo temperature $\TK$ is a new, many-body energy
scale that emerges for a singly-occupied Anderson impurity
\cite{hewson}. It is essentially the binding energy of the spin
singlet formed between the localized, unpaired electron and electrons
in the surrounding reservoirs; $k\TK \approx 4 - 250\,\mu$eV in our
SET, depending on the other tunable parameters.

The conductance $G$ of an SET is analogous to the resistivity $\rho$
of a bulk Kondo system.  Although one thinks of the increase in
resistivity at low $T$ as the hallmark of the Kondo effect, transport
properties have proven more difficult to calculate than thermodynamic
properties. For $T \ll \TK$, $\rho$ is theoretically and
experimentally known to equal $\rho_0 - c T^2$ (Fermi liquid behavior)
\cite{nozieres} and for $\TK < T < 10\,\TK,\ \rho$ is roughly
logarithmic in $T$ \cite{kondo,hamann}, but the crossover region has
only recently been successfully treated \cite{costi}.

Furthermore, the Anderson model has several interesting regimes
parametrized by $\tilde\epsilon_0 \equiv \epsilon_0/\Gamma$: the Kondo
regime $\tilde\epsilon_0 \ll -0.5$, the mixed-valence regime
$-0.5\,\Lap\,\tilde\epsilon_0\,\Lap\,0$, and the empty orbital regime
$\tilde\epsilon_0\,\Gap\,0$, each of which has different transport
properties. The Kondo regime describes many systems of dilute magnetic
impurities in metals, while the mixed-valence regime provides some
understanding of heavy-fermion compounds
\cite{mixedv,Varma,bickers85}.  We know of no material described by
the empty orbital regime.  Though conductance through an SET
normalized to its zero-temperature value $\tilde G(\tilde T) \equiv
G(T/\TK)/G_0$ is expected to be universal in the Kondo regime, where
the only small energy scale is $\TK$, it should change as
$\tilde\epsilon_0 \rightarrow 0$ (the mixed-valence regime), where
$\TK$ and $\Gamma$ become comparable \cite{costi}. The great advantage
of the SET is that $\epsilon_0$ can be tuned by varying $\Vg$ to test
the predictions for all regimes in one and the same system.

As $\Vg$ is varied, the conductance of an SET undergoes
oscillations caused by what is usually called the Coulomb
blockade. Current flow is possible in this picture only when two
charge states of the droplet are degenerate, i.e. $\epsilon_0 =0$ or
$\epsilon_0 + U = 0$, marked by vertical dashed lines in Fig.\
\ref{cbfig} as determined by the analysis of Fig.\ \ref{paramfit}.
The conductance between these dashed lines is expected to be very
small. However, in this range the charge state of the site is odd, as
portrayed in Fig.\ \protect\ref{energydiagram}, and the Kondo effect
allows additional current flow. Strikingly, at low temperature (dots,
100~mK and triangles, 800~mK), the conductance maxima do not even
occur at $\epsilon_0 = 0$ and $\epsilon_0 + U = 0$ --- the Kondo
effect makes the off-resonant conductance even larger than the
conductance at the charge-degeneracy point
\cite{wingreen.kondo}. Raising the temperature suppresses the Kondo
effect, causing the peaks to approach the positions of the bare
resonances.

The inset of Fig.\ \ref{cbfig}
shows how $\Gamma$ is determined: For $T \ge \Gamma/2$, the
conductance peak is well-described by the convolution of a
Lorentzian of FWHM $\Gamma$ with the derivative of a Fermi-Dirac
function (FWHM $3.52 kT$). This convolution has a FWHM $0.78 \Gamma +
3.52 kT$, so extrapolating the experimentally-measured linear
dependence back to $T=0$ gives $\Gamma = 295\pm20\,\mu$eV.

When the energy of the localized state is far below the Fermi level
 ($\tilde \epsilon_0 \ll -1$), scaling theory predicts that $\TK$
 depends exponentially on the depth of that level \cite{haldane}:
\begin{equation}
\TK = \frac{\sqrt{\Gamma U}}{2} e^{\pi \epsilon_0 (\epsilon_0 +
U)/\Gamma U}.
\label{TK}
\end{equation}
Note that, because $U$ is finite, $\log \TK$ is quadratic in
$\epsilon_0$. This strong dependence on $\epsilon_0$ causes the
Kondo-enhanced conductance to persist to higher temperatures near
$\epsilon_0 = 0$ (and near $\epsilon_0 = -U$, by particle-hole
symmetry) than in-between. In fact, at $T=0$ the conductance should
sustain its maximum value all the way between the two observed peaks
in Fig.\ \protect\ref{cbfig}
\cite{nglee.kondo,glazmanraikh.kondo,gg.kondo} (see Fig.\
\protect\ref{paramfit}(b) for expected $G(\tilde\epsilon_0)$ at
$T=0$), but in the valley even our $\Tb \simeq 100$ mK $\ >\ \TK
\approx 40$ mK.

Figure\ \ref{tdepfig}(a) shows that, for fixed
$\tilde\epsilon_0$ in the Kondo regime, $G \sim -\log(T)$ over as much
as an order of magnitude in temperature, beginning at $\Tb$. Thermal
fluctuations in localized state occupancy cut off the $\log(T)$
conductance for $kT\,\Gap\,|\epsilon_0|/4$, consistent with
simulations of thermally-broadened Lorentzian resonances. As
$\tilde\epsilon_0 \rightarrow 0$ (Fig.\ \ref{tdepfig}(b)), $\TK$
increases, as evinced by the saturation of the conductance at low
temperature.

To fit the experimental data for each $\epsilon_0$ we use the empirical form
\begin{equation}
G(T) = G_0 \Big(\frac{\Tp2}{T^2 + \Tp2}\Big)^{\textstyle s},
\label{GofT}
\end{equation}
where $\Tpr$ is taken to equal $\TK/ \sqrt{2^{1/s}-1}$ so that $G(\TK)
= G_0/2$. For the appropriate choice of $s$, which determines the
steepness of the conductance drop with increasing temperature, this
form provides a good fit to numerical renormalization group results
\cite{costi} for the Kondo, mixed-valence, and empty orbital regimes,
giving the correct Kondo temperature in each case. The parameter $s$
is left unconstrained in the fit to our data, but its fit value is
nearly constant at $0.20 \pm 0.01$ in the Kondo regime, while as
expected it varies rapidly as we approach the mixed-valence regime
(Figure~\ref{tdepfig}(b)).  The expected value of $s$ in the Kondo
regime depends on the spin of the impurity: $s=0.22 \pm 0.01$ for
$\sigma=1/2$.

Using the values of $G_0$ and $\TK$ extracted in this way we confirm
that $\tilde G$ is universal in the Kondo regime. Figure\
\ref{scaledfig} shows $\tilde G(\tilde T)$ for data like those of
Figure\ \ref{tdepfig} for various values of $\tilde\epsilon_0 \sim -1$
(on the left peaks of Fig.\ \ref{cbfig}). We have also included data
from the same SET, but with $\Gamma$ reduced by $25 \%$ by adjusting
the point-contact voltages.  The data agree well with numerical
renormalization group calculations by Costi and Hewson (solid line)
\cite{costi}. In the mixed-valence regime it is
difficult to make a quantitative comparison between theoretical
predictions and our experiment. Qualitatively, in both calculation and
experiment, $\tilde G(\tilde T)$ exhibits a sharper crossover between
constant conductance at low temperature and logarithmic dependence at
higher temperature in the mixed-valence regime than in the Kondo
regime (see Fig.\ \ref{scaledfig}) \cite{costi}.

In Figure\ \ref{paramfit}(a), we plot
$\TK(\tilde\epsilon_0)$ extracted from our fits, along with the
theoretical prediction (Eq.~\ref{TK}) for the Kondo regime. The value
of $\Gamma = 280\pm10\,\mu$eV extracted is in good agreement with the
value $\Gamma = 295\pm20\,\mu$eV determined as illustrated in Fig.\
\ref{cbfig} (inset). The prefactor is approximately three times larger
than $\sqrt{\Gamma U}/2$, which must be considered good agreement
given the simplifying assumptions in the calculations and the
sensitivity to the value of the exponent \cite{multilev}.

$G_0$ is predicted to vary with the site occupancy $\nd$, and hence
also with $\tilde\epsilon_0$, according to the Friedel sum rule
\begin{equation}
G_0(\nd) = G_{\mbox{\footnotesize max}}\sin^2\Big(\frac{\pi}{2} \nd\Big),
\label{friedel}
\end{equation}
where $G_{\mbox{\protect\footnotesize max}}$ is the unitary limit of
transmission: $2 e^2/h$ if the two barriers are symmetric, less if
they are asymmetric. For small $|\tilde\epsilon_0|$, $\TK \gg \Tb$, so
we can directly measure the value of $G_0$. Even when $\TK$ is not
$\gg \Tb$, we can still extract the value of $G_0$ from our fit.  In
Figure\ \ref{paramfit}(b), we compare the combined results of both
these methods with $G_0(\tilde\epsilon_0)$ inferred from a
non-crossing approximation (NCA) calculation \cite{wingreen.kondo} of
$\nd(\tilde\epsilon_0)$ according to Equation~\ref{friedel}. The
agreement is excellent except outside the left peak, where
experimentally the conductance does not go to zero even at zero
temperature (see Fig.\ \ref{cbfig}) \cite{discrepancy}.

We have demonstrated quantitative agreement between transport
measurements on an SET and calculations for a spin-$1/2$ Anderson
impurity. The SET allows us to both accurately
measure and vary $\Gamma$ and $\epsilon_0$, and to observe their effect
on $\TK$ and $G_0$. We have also observed the cross-over between the
Kondo and mixed-valence regimes.

We acknowledge fruitful discussions with David Abusch-Magder, Igor
Aleiner, Ray Ashoori, Gene Bickers, Daniel Cox, Leonid Glazman,
Selman Hershfield, Wataru Izumida, Steven Kivelson, Leo Kouwenhoven,
Patrick Lee, Leonid Levitov, Avraham Schiller, Chandra Varma, and
especially Ned Wingreen, Yigal Meir, and Theo Costi. Theo Costi,
Wataru Izumida, and Ned Wingreen generously provided data from their
prior calculations. D. G.-G. thanks the Hertz Foundation, and
J. G. thanks NEC, for graduate fellowship support. This work was
supported by the US Army Research Office Joint Services Electronics
Program under contract DAAG 55-98-1-0080, by the US Army Research
Office under contract DAAG 55-98-1-0138, and by the MRSEC Program of
the National Science Foundation under Award No. DMR94-00334.



\begin{thebibliography}{10}
\bibitem{note.kondo} We have now observed the Kondo effect in four
SETs. All the data presented in this Letter are from a single SET
nominally identical to, but distinct from, that discussed in
\protect\cite{gg.kondo}.
\bibitem{mixedv} In the latter case, a single-impurity
picture is incomplete: inter-impurity interactions must also play a major 
role.
\bibitem{discrepancy} There is no zero-bias peak in differential
conductance in this region, demonstrating that the extra conductance
is {\em not} caused by the Kondo effect here.
\bibitem{valleyfit} When $-\tilde\epsilon_0\,\Gap\,2$ or
$-\tilde\epsilon_0\,\Lap\,5.5$, the measured $\TK$ and $G_0$ sharply
decrease below their predicted values. This could be caused by an
unintentional ac bias $\Vac$ applied across the SET. Such a bias has
often been known to raise the effective temperature of small electron
systems. Here, it could have an even more drastic effect on the Kondo
resonance, and hence on the conductance, at these values of
$\tilde\epsilon$, for which $\Vac\,\Gap\,\TK$.
\end{thebibliography}

\newcommand{\noopsort}[1]{} \newcommand{\printfirst}[2]{#1}
  \newcommand{\singleletter}[1]{#1} \newcommand{\switchargs}[2]{#2#1}

Figures printed separately:

\bef[htb]
\caption{}
\label{energydiagram}
\enf

\bef[htb]
\caption{}
\label{cbfig}
\enf

\bef[htb]
\caption{}
\label{tdepfig}
\enf

\bef[htb]
\caption{}
\label{scaledfig}
\enf

\bef[htb]
\caption{}
\label{paramfit}
\enf

\end{multicols}

\end{document}